\documentclass[preprint,12pt]{elsarticle}

 \usepackage{graphicx}
\usepackage{subfigure}

\usepackage{amssymb}

\journal{Physics Letters A}

\begin{document}

\begin{frontmatter}



\title{Capacitance and charging of metallic objects}


\author{Titus Sandu}
 \ead{titus.sandu@imt.ro}
\ead{tel:+40-21-269.07.70/ext.30}
\ead{fax:+40-21-269.07.72}

\address{National Institute for Research and Development in Microtechnologies-IMT,
126A, Erou Iancu Nicolae Street, 077190, Bucharest, ROMANIA}

\author{George Boldeiu}

\address{National Institute for Research and Development in Microtechnologies-IMT,
126A, Erou Iancu Nicolae Street, 077190, Bucharest, ROMANIA}

\author{Victor Moagar-Poladian}

\address{National Institute for Research and Development in Microtechnologies-IMT,
126A, Erou Iancu Nicolae Street, 077190, Bucharest, ROMANIA}

\begin{abstract}
The capacitance of arbitrarily shaped objects is reformulated in terms of the Neumann-Poincar\'{e} operator. Capacitance is simply the dielectric permittivity of the surrounding medium multiplied by the area of the object and divided by the squared norm of the Neumann-Poincar\'{e} eigenfunction that corresponds to its largest eigenvalue. The norm of this eigenfunction varies slowly with shape changes and allows perturbative calculations. This result is also extended to capacitors. For axisymmetric geometries a numerical method provides excellent results against finite element method results. Two scale-invariant shape factors and the capacitance of nanowires and of membrane in biological cells are discussed.   
\end{abstract}

\begin{keyword}
Capacitance \sep Laplace equation \sep Neumann-Poincar\'{e} operator \sep Boundary intergral equation \sep Localized  surface plasmon resonances \sep 
Dielectric spectra of live cells

\PACS 41.20.Cv \sep 82.47.Uv \sep 87.19.rf \sep 87.50.C-


\end{keyword}

\end{frontmatter}


\section{Introduction}

Since the 19th century potential theory has been growing continuously as 
gravitational and electromagnetic interactions and forces were derived 
from potentials satisfying Laplace, Poisson and Helmholtz equations. Some 
boundary value problems such as the Dirichlet problem and the Neumann 
problem, the electrostatic distribution of charges on conductors or the 
Robin problem can all be defined in terms of potential theory. For domains 
with sufficiently smooth boundaries the above problems use specific types of 
potentials like the volume potential, the single- and the double-layer potentials, 
the logarithmic potential for two-dimensional domains, etc. \cite{Kellog1967,Khavinson2007}. 

The Dirichlet and Neumann problems defined on domains with sufficiently 
smooth boundaries (i.e., a regular piecewise Lyapunov surface \cite{Gunter1934}) can be 
recast in integral equations which lead to compact operators on domain 
boundary: the Neumann-Poincar\'{e} operator and its adjoint \cite{Kellog1967,Khavinson2007}. These 
methods are successfully applied in some practical and physical problems 
regarding dielectric heterogeneous systems \cite{Bergman1978} like 
radio-frequency and microwave dielectric spectra of living cells \cite{Vrinceanu1996} and 
plasmonic properties of metallic nanoparticles \cite{Ouyang1989,Mayergoyz2005,Davis2009}. 

Another problem which stems from potential theory is the equilibrium charge 
distribution on a conductor (the Robin problem) \cite{Robin1886} and the implicit 
capacitance with applications in computational biophysics \cite{Simonson2003}, in scanning 
probe microscopy \cite{Hofer2003,Hudlet1998}, or in electrical charge storage in supercapacitors 
\cite{Simon2008}. The capacitance of an arbitrarily shaped body is directly related to 
hydrodynamic friction and to diffusion-controlled reaction rate \cite{Hubbard1993}. Thus a 
method of calculating the capacitance is based on ``mimicking'' the 
diffusion-controlled reaction rate on an object of arbitrary shape as random 
walks \cite{Douglas1994}. Another random walk method is defined just on the boundary 
and is given by the ergodic generation of the equilibrium charge 
distribution \cite{Mascagnia2004}. 

A standard procedure for solving the Laplace equation is the Finite Element 
Method (FEM) \cite{Johnson1987} and the Boundary Integral Equation (BIE) method which 
leads to a finite element formulation as the Boundary Element Method (BEM) 
\cite{Poljak2005}. In the FEM the entire domain is discretized by using 
elements and the associated basis functions, while in the BEM only 
the surfaces of the inclusions within the domain are used for discretization. 
The bounded and the unbounded domains are on equal 
footing in the BEM. On the other hand, the FEM solves the static and 
quasi-static electromagnetic field problems in unbounded domains by using 
one of the additional ingredients: infinite elements \cite{Bettess1977}, coordinate transformations \cite{Lowther1989}, 
or hybrid FEM/BEM methods \cite{Salon1988}. Infinite domains can be elegantly and 
efficiently treated with the hybrid Trefftz method, where, beside the usual 
elements, an additional set of functions is employed to treat 
singularities and infinities \cite{Qin2005}.

In this paper, by a BIE method \cite{Sandu2010,Sandu2011,Sandu2013}, we will define the capacitance and the equilibrium charge of an 
arbitrarily shaped object in terms of the eigenvectors of the 
Neumann-Poincar\'{e} operator and of its adjoint \cite{Kellog1967,Khavinson2007}. Compared to other methods the operator based BIE method 
provides directly the geometric dependence of 
capacitance that is incorporated in an eigenvector norm of the Neumann-Poincar\'{e} operator. This eigenvector corresponds to the largest eigenvalue. 
Accordingly, capacitance is obtained concurrently with other physical properties like localized plasmon resonances of metallic 
nanoparticles that have the same shape \cite{Ouyang1989,Mayergoyz2005,Davis2009}. Furthermore, operator perturbations on 
eigenvalues and on eigenvectors permit the perturbative estimation of capacitance and other physical properties.
The reformulation of capacitance allows defining several scale-invariant shape factors that can also be readily used in the estimation of 
capacitance for arbitrary shapes. Applications regarding the capacitance of a general capacitor, the membrane capacitance of living cells, charge 
storage in supercapacitors, and the electric capacitance of molecular nanowires are discussed. 

The paper is organized as follows. The method is described in the second section. Then, the connection with a general capacitor 
is made in the following section. 
Section 4 describes the numerical method, its accuracy and some applications. The conclusions are summarized in the last section.

\section{The operator approach in potential theory. Capacitance of a metallic object}

We assume an arbitrarily shaped domain $\Omega $ of volume $V$ and bounded by 
the surface $\Sigma $. We denote the complementary set of $\Omega $ by 
$\Omega ^c\mbox{ = }\Re ^3\backslash \bar {\Omega }$, where $\bar {\Omega }$ 
is the closure of $\Omega $ and $\Re ^3$ is the Euclidian 3-dimensional 
space. The following operators can be defined on $\Sigma $:

\begin{equation}
\label{eq1}
\hat {M}\left[ u \right] = \frac{1}{4\pi }\int\limits_{\bf{x},\bf{y} \in \Sigma } 
{\frac{u\left( \bf{y} \right)\bf{n}\left( \bf{x} \right) \cdot \left( {\bf{x} - \bf{y}} 
\right)}{\left| {\bf{x} -\bf{ y}} \right|^3}d\Sigma \left( \bf{y} \right)} ,
\end{equation}

\noindent
its adjoint,

\begin{equation}
\label{eq2}
\hat {M}^\dag \left[ v \right] = \frac{1}{4\pi }\int\limits_{\bf{x},\bf{y} \in \Sigma 
} {\frac{v\left( \bf{y} \right)\bf{n}\left( \bf{y} \right) \cdot \left( {\bf{x} - \bf{y}} 
\right)}{\left| {\bf{x} - \bf{y}} \right|^3}d\Sigma \left( \bf{y} \right)} ,
\end{equation}

\noindent
which is the Neumann-Poincar\'{e} operator \cite{{Khavinson2007}}, and the symmetric and non-negative Coulomb operator

\begin{equation}
\label{eq3}
\hat {S}\left[ u \right] = \frac{1}{4\pi }\int\limits_{\bf{x},\bf{y} \in \Sigma } 
{\frac{u\left( \bf{y} \right)}{\left| {\bf{x} - \bf{y}} \right|}d\Sigma \left( \bf{y} \right)} 
.
\end{equation}

\noindent
In Eqs. (\ref{eq1})-(\ref{eq2}) \textbf{n} is the normal vector to $\Sigma $. The 
operator $\hat {M}$ can be made symmetric via the 
Plemelj's symmetrization principle $\hat {M}^\dag \,\hat {S} = \hat 
{S}\,\hat {M}$ \cite{Khavinson2007}. Thus, the operator $\hat {M}$ is symmetric with respect 
to the inner product defined by the symmetric 
and non-negative operator $\hat {S}$

\begin{equation}
\label{eq4}
\left\langle {v} | {u} \right\rangle _S = \left\langle {v} | {\hat {S}\left[ u \right]} \right\rangle ,
\end{equation}

\noindent
where $\left\langle {} \mathrel{\left| 
{\vphantom { }} \right. \kern-\nulldelimiterspace} {} \right\rangle$ defines the standard inner 
product on $L^2\left( \Sigma \right)$ and $\left\langle {} \mathrel{\left| 
{\vphantom { }} \right. \kern-\nulldelimiterspace} {} \right\rangle _S $ is 
the inner product determined by $\hat {S}$. The Hilbert space $L^2\left( \Sigma \right)$ is the vector space 
of all square-integrable functions defined on $\Sigma$. The standard inner product on $L^2\left( \Sigma \right)$ 
of two functions $u_1 \left( \bf{x} \right)$ and $u_2 \left( \bf{x} \right)$ is defined as

\begin{equation}
\label{eq5}
\left\langle {u_1 } \mathrel{\left| {\vphantom {{u_1 } {u_2 }}} \right. 
\kern-\nulldelimiterspace} {u_2 } \right\rangle = \int\limits_{\bf{x} \in \Sigma 
} {u_1^\ast \left( x \right)\,u_2 \left(\bf{x} \right)d\Sigma \left( \bf{x} \right)} 
,
\end{equation}

\noindent
where the star sign * signifies the complex conjugate of a complex number. 
The operators $\hat {M}$ and $\hat {M}^\dag $ have the same spectrum bounded 
by the interval [-1/2 ,1/2] and the eigenfunctions $u_i $ of $\hat {M}$ are 
related to the eigenfunctions $v_i $ of $\hat {M}^\dag $ by $v_i = \hat 
{S}\left[ {u_i } \right]$, which makes them bi-orthogonal, i.e. if $\hat 
{M}\left[ {u_i } \right] = \chi _i u_i $ and $\hat {M}^\dag \left[ {v_j } 
\right] = \chi _j v_j $, then $\left\langle {v_j } \mathrel{\left| 
{\vphantom {{v_j } {u_i }}} \right. \kern-\nulldelimiterspace} {u_i } 
\right\rangle = \delta _{ij} $ \cite{Ouyang1989, Khavinson2007, Sandu2010}. 
In physical terms $\hat {M}\left[ u \right]$ 
is related to the normal component of the electric field that is generated 
by surface charge density $u$. On the other hand, $v_i $ is the electric potential 
generated on surface $\Sigma $ by the charge distribution $u_i $. The 
largest eigenvalue of $\hat {M}$ and $\hat {M}^\dag $ is 1/2 irrespective of 
the domain shape and its corresponding eigenfunction $v_1 $ of $\hat 
{M}^\dag $ is a constant function, i. e., $v_1 = $constant on $\Sigma $, 
which comes from the following relation \cite{Khavinson2007}

\begin{equation}
\label{eq6}
\hat {M}^\dag \left[ {v_1 } \right] = \frac{\mbox{constant}}{4\pi 
}\int\limits_{\bf{y} \in \Sigma } {\frac{\bf{n}(\bf{y}) \cdot \left( {\bf{x} - \bf{y}} \right)}{\left| 
{\bf{x} - \bf{y}} \right|^3}d\Sigma \left( \bf{y} \right) = \frac{1}{2}} \,v_1. 
\end{equation}
Equation (\ref{eq6}) can be interpreted in terms of the solid angle under which 
$\Sigma $ can be seen from an arbitrary point located also on $\Sigma $. As 
we will see below, the companion eigenfunction $u_1 $ of $\hat {M}$ is 
proportional to the equilibrium charge distribution on a conductor of shape 
determined by $\Sigma $. We note 
that the spectrum of $\hat {M}$ and $\hat {M}^\dag $ is invariant under the 
transformation ${\bf{x}} \to t {\bf{x}}$, $t > 0$ (the scale invariance). In contrast, the 
spectrum of $\hat {S}$ is proportional to the linear size of $\Omega $. This 
can be easily checked in the case of a sphere where the eigenvalues of $\hat 
{S}$ are proportional to the radius of the sphere. Thus, because of $\hat 
{S}$, $u_i $ and $v_i $ are not scale invariant. Therefore, the norms of 
$u_i $ and $v_i $ are proportional to the square-root and to the inverse of the 
square-root of the linear size of $\Omega $, respectively. The size-dependence 
of $u_i $ and $v_i $ is also reflected in the the bi-orthogonality 
conditions defined in Ref. \cite{Ouyang1989}. 
Operators (\ref{eq1})-(\ref{eq3}) can be used in the resolution of many physical problems. 
Dielectric spectra of living cells in microwave and radio-frequency regimes 
\cite{Vrinceanu1996, Sandu2010} or plasmonic properties of metallic nanoparticles 
\cite{Ouyang1989,Mayergoyz2005,Davis2009, Sandu2011,Sandu2013} can be treated 
with the help of $\hat {M}$ and $\hat {M}^\dag $ whose inversion is carried out via 
the eigenfunctions $u_i $ and $v_i $ and the corresponding  eigenvalues $\chi_i$. However, 
the above problems do not require normalized eigenfunctions because $u_i $ and $v_i $ come in pairs in the 
desired solutions \cite{Vrinceanu1996,Ouyang1989,Mayergoyz2005,Davis2009, Sandu2010,Sandu2011,Sandu2013}. 
Thus, there is no need for 
explicit calculation of $\hat {S}$. Operator $\hat {S}$ is used whenever 
an explicit expression for the near-field properties like the near-field 
enhancement produced by localized plasmon resonances are needed \cite{Sandu2013}. 
Another example where  $\hat {S}$ would be needed is the calculation of capacitance and of equilibrium charge on a 
metallic object. 
The Robin problem of finding the equilibrium charge distribution on a 
conductor of arbitrary shape can be 
cast into the operator form as \cite{Robin1886}

\begin{equation}
\label{eq19}
\hat {M}\left[{u_R } \right] = \frac{1}{2}u_R \left( {\bf{x}} \right),
\end{equation}

\noindent
with the constraint $\int\limits_{{\bf{x}} \in \Sigma } {u_R \left( {\bf{x}} 
\right)d\Sigma  \left( {\bf{x}} \right) } = 1$. The constraint (normalization condition) can be put in the following form

\begin{equation}
\label{eq20}
\left\langle {1\left| {u_R } \right\rangle = 1} \right.,
\end{equation}

\noindent
where $1 \in L^2\left( \Sigma \right)$ is the constant distribution that has 
the value 1 on $\Sigma $. Equation (\ref{eq19}) is found from the jump formula 
obeyed by the derivative of the single-layer potential across $\Sigma $ \cite{Khavinson2007} 
and it has the obvious solution $u_R \propto u_1 $. In other words, the 
solution of Robin problem is proportional to the eigenfunction of the 
largest eigenvalue of $\hat {M}$. The constant value $V_R $ of the electric 
potential generated by $u_R $ is formally given by

\begin{equation}
\label{eq21}
\hat {S}\left[ {u_R } \right] = V_R 1
\end{equation}

\noindent
and it is called the Robin constant, while its inverse is the capacitance C of 
the body bounded by $\Sigma $ \cite{Kellog1967}. The left-side hand 
of Eq. (\ref{eq21}) needs to be divided by dielectric permittivity $\varepsilon $ of 
the embedding medium if any physical situation is considered. Thus, 
from (\ref{eq20}) and (\ref{eq21}), in the general case of an outer medium with dielectric 
permittivity $\varepsilon $, the capacitance is:

\begin{equation}
\label{eq22}
C = \frac{1}{V_R } = \frac{\varepsilon }{\left\langle {u_R \left| {\hat 
{S}\left[ {u_R } \right]} \right\rangle } \right.}.
\end{equation}

\noindent
The quantity $W_{u_R } = {\left\langle {u_R \left| {\hat {S}\left[ {u_R } 
\right]} \right\rangle } \right.} \mathord{\left/ {\vphantom {{\left\langle 
{u_R \left| {\hat {S}\left[ {u_R } \right]} \right\rangle } \right.} {\left( 
{2\varepsilon } \right)}}} \right. \kern-\nulldelimiterspace} {\left( 
{2\varepsilon } \right)}$ is just the electrostatic energy of the charged 
metallic body with charge density $u_R $. The electrostatic energy $W_{u_R }$ of the 
equilibrium charge density $u_R $ is the energy minimum attained for the set 
of arbitrary surface charge densities that satisfy the constraint provided by 
Eq. (\ref{eq20}). This is the Thomson theorem \cite{Landau1984} and it is used as the 
definition for the equilibrium charge distribution in abstract mathematical terms \cite{Kellog1967}. 
The proof is as follows. For any surface density $u$ one has the 
expansion $u = \sum\limits_{i = 1}^\infty {a_i u_i } $, where $a_i $ are the 
expansion coefficients of the surface density $u$. In addition, $u$ obeys 
$\left\langle {1\left| u \right\rangle = 1} \right.$, which fixes the first 
expansion coefficient since $\left\langle {1\left| {u_i } 
\right\rangle = 0} \right.$ for $i \ne 1$ (the biorthogonality condition). 
One can prove that the constant $a_1 $ is the proportionality factor between $v_1 $ and 
the constant distribution $1$, i. e., $1 = {v_1 } \mathord{\left/ 
{\vphantom {{v_1 } {a_1 }}} \right. \kern-\nulldelimiterspace} {a_1 }$. Thus 

\begin{equation}
\label{eq23}
\begin{array}{l}
 \left\langle {u\left| {\hat {S}\left[ u \right]} \right\rangle } \right. = 
\sum\limits_{i,j = 1}^\infty {\left\langle {a_i^\ast u_i \left| {\hat 
{S}\left[ {a_j u_j } \right]} \right\rangle } \right.}  = \\ 
 =\sum\limits_{i,j = 1}^\infty {a_i^\ast a_j \left\langle {u_i \left| {v_j } 
\right\rangle } \right.} = \sum\limits_{i,j = 1}^\infty {a_i^\ast a_j \delta 
_{ij} } = \\ 
= \sum\limits_{i = 1}^\infty {\left| {a_i } \right|^2} \ge \left| 
{a_1 } \right|^2  = \left\langle {u_R \left| {\hat {S}\left[ {u_R } \right]} \right \rangle } . \right .
 \end{array}
\end{equation}

\noindent
which demonstrates the Thomson theorem. Moreover, the explicit expression of 
the equilibrium distribution is 

\begin{equation}
\label{ur}
u_R = a_1 u_1 
\end{equation}
and $a_1 $ is related to the norm of $v_1 $ by

\begin{equation}
\label{eq24}
\left\langle {v_1 \left| {v_1 } \right\rangle } \right. = \left\| {v_1 } 
\right\|^2 = a_1^2 \left\langle {1\left| 1 \right\rangle } \right. = a_1^2 
A,
\end{equation}

\noindent
where $A$ is the area of $\Sigma $. Finally, considering  Eqs. (\ref{ur}) and (\ref{eq24}) the capacitance 
expression (\ref{eq22}) takes a simple and compact form

\begin{equation}
\label{eq25}
C = \frac{\varepsilon A}{\left\| {v_1 } \right\|^2}.
\end{equation}

\noindent
We have pointed out earlier that $\left\| {v_1 } \right\|^2$ is proportional to the linear size of 
$\Omega $, therefore the capacitance itself is also proportional to the linear size of 
the body. Equation (\ref{eq25}) shows explicitly both the geometric dependence of 
capacitance of an arbitrarily shaped object and the scale invariance of the shape factor $C / \sqrt{4 \pi A }$. 
The shape factor varies slowly with the conductor shape \cite{Chow1982} 
thus, one can say that $\left\| {v_1 } \right\|^2$ is a slowly varying 
function of the conductor shape.
In some cases like that of a cube \cite{Hwang2004} the capacitance is difficult to compute, but as it is suggested by 
Eq. (\ref{eq25}) operator perturbations can be used to estimate the capacitance of any object \cite{Grieser2009}.  

\begin{figure}
 \begin{center}
\includegraphics [width=3 in] {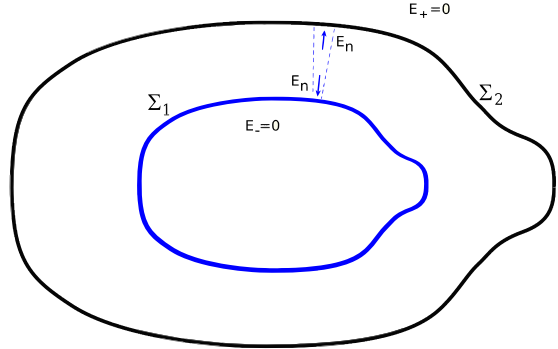}
  \end{center}
\caption{\label{fig:1}
(Color online) Schematic representation of a capacitor with the configurations of fields inside $\Sigma _1 $ and 
outside  $\Sigma _2 $. The dotted lines delimitate a Gaussian surface that is used further in the main text by invoking the Gauss law.}
\end{figure}

\section{Capacitors and their capacitance}

The capacitance of an arbitrary object is defined and calculated with respect to 
infinity where the electric potential is supposed to vanish. In general, a capacitor 
consists of two separated conducting bodies. We consider a 
capacitor that is made of two closed and smooth surfaces $\Sigma _1 $ and $\Sigma _2 
$ in which $\Sigma _2 $ encloses $\Sigma _1 $ (Fig. \ref{fig:1}). The capacitor 
capacitance is defined as the total charge that is held on $\Sigma _1 $ when 
the electrical potential is 1 on $\Sigma _1 $ and 0 on $\Sigma _2 $. This 
definition can be cast in a form in which the total charge on $\Sigma _1 $ 
is given by the Gauss law or by the total electrostatic energy 
enclosed within the space between $\Sigma _1 $ and $\Sigma _2 $ \cite{Landau1984}. The 
electrostatic problem of a capacitor is to find a function that is 1 on 
$\Sigma _1 $ and 0 on $\Sigma _2 $ and obeys the Laplace equation in the 
space $\Omega _{12} $ between $\Sigma _1 $ and $\Sigma _2 $:

\begin{equation}
\label{eq26}
\begin{array}{l}
 \Delta u\left( {\bf{x}} \right) = 0;\;{\bf{x}} \in \Omega _{12} \\ 
 u\left( {\bf{x}} \right) = 1;\;{\bf{x}} \in \Sigma _1 \\ 
 u\left( {\bf{x}} \right) = 0;\;{\bf{x}} \in \Sigma _2 .\\ 
 \end{array}
\end{equation}

\noindent
It is easy to see that, with the boundary condition taken from (\ref{eq26}), the solution of the Laplace equation 
inside of $\Sigma _1 $ and outside of $\Sigma _2 $ is the constant 1 and the 
constant 0, respectively. Inside $\Omega _{12} $ we seek a 
solution for (\ref{eq26}) in the form of two single-layer 
potentials

\begin{equation}
\label{eq27}
u\left( x \right) = \frac{1}{4\pi }\int\limits_{
 {\bf{y}} \in \Sigma _1  } {\frac{\mu _1 \left( {\bf{y}} \right)}{\left| {\bf{x}} - {\bf{y}} \right|}d\Sigma 
\left( {\bf{y}} \right)} + \frac{1}{4\pi }\int\limits_{
 {\bf{y}} \in \Sigma _2 } {\frac{\mu _2 \left( {\bf{y}} \right)}{\left| {\bf{x}} - {\bf{y}} \right|}d\Sigma 
\left( {\bf{y}} \right)} ,
\end{equation}

\noindent
where $\mu _1 $ and $\mu _2 $ are the induced charge densities on $\Sigma _1 
$ and $\Sigma _2 $. Similar to (\ref{eq1}) we define on $\Sigma _1 $ and $\Sigma _2 
$ four operators $\hat {M}_{11} $, $\hat {M}_{12} $, $\hat {M}_{21} $, and 
$\hat {M}_{22} $ as

\begin{equation}
\label{eq28}
\hat {M}_{ij} \left[ {\mu _j } \right] = \frac{1}{4\pi 
}\int\limits_{
 {\bf{x}} \in \Sigma _i  
 {\bf{y}} \in \Sigma _j  
 } {\frac{\mu _j \left( \bf{y} \right) \bf{n}\left( \bf{x} \right) \cdot \left( 
{\bf{x} - \bf{y}} \right)}{\left| {\bf{x} - \bf{y}} \right|^3}d\Sigma \left( \bf{y} \right)}, 
\end{equation}

\noindent
with $i,j = \overline {1,2} $. The equations obeyed by $\mu _1 $ and $\mu _2 
$ are

\begin{equation}
\label{eq29}
\begin{array}{l}
 \hat {M}_{11} \left[ {\mu _1 } \right] + \hat {M}_{12} \left[ {\mu _2 } 
\right] = \frac{1}{2}\mu _1 \\ 
 \hat {M}_{21} \left[ {\mu _1 } \right] + \hat {M}_{22} \left[ {\mu _2 } 
\right] = - \frac{1}{2}\mu _2, \\ 
 \end{array}
\end{equation}

\noindent
which say that the normal fields on $\Sigma _1 $ from inside and on $\Sigma _2 $ from 
outside are zero. The solution of (\ref{eq29}) is the solution of 
(\ref{eq26}) up to a multiplicative constant. The multiplicative constant is fixed 
by the equations that set the boundary conditions of (\ref{eq26}):

\begin{equation}
\label{eq30}
\begin{array}{l}
 \hat {S}_{11} \left[ {\mu _1 } \right] + \hat {S}_{12} \left[ {\mu _2 } 
\right] = 1 \\ 
 \hat {S}_{21} \left[ {\mu _1 } \right] + \hat {S}_{22} \left[ {\mu _2 } 
\right] = 0, \\ 
 \end{array}
\end{equation}

\noindent
with $\hat {S}_{ij} $ being similar to (\ref{eq3})

\begin{equation}
\label{eq31}
\hat {S}_{ij} \left[ u \right] = \frac{1}{4\pi 
}\int\limits_{
 {\bf{x}} \in \Sigma _i   
 {\bf{y}} \in \Sigma _j  
 } {\frac{u\left( {\bf{y}} \right)}{\left| {\bf{x}} - {\bf{y}} \right|}d\Sigma 
\left( {\bf{y}} \right)} .
\end{equation}

The capacitance of the capacitor is the total charge represented by charge 
density $\mu _1 $ and depends not only on operators like (\ref{eq1}) and (\ref{eq2}) but also on 
inter-surface operators (\ref{eq28}). 
In the special case when $\Sigma _2 $ 
is an equipotential surface determined by the equilibrium charge distributed 
on $\Sigma _1 $ a compact formula based on operators (\ref{eq1})-(\ref{eq2}) can be deduced 
for capacitance. It is not hard to see that Eqs. (\ref{eq29}) have as 
solutions the charge densities $\mu _1 $ and $\mu _2 $, which are proportional to the equilibrium 
charge densities induced on $\Sigma _1 $ and $\Sigma _2 $, respectively. In addition, $\mu _1 $ and 
$\mu _2 $ must have opposite signs. To determine $\mu _1 $ and $\mu _2 $, and the 
capacitor capacitance one needs also Eqs. (\ref{eq30}). Thus, by integrating the first 
equation of (\ref{eq30}) on $\Sigma _1 $ and the second equation on $\Sigma _2 $ one obtains the following relations

\begin{equation}
\label{eq32}
\begin{array}{l}
V_1 + V_2 = 1 \\
V_{12} + V_2 = 0 \\
\end{array}
\end{equation}

\noindent
where $V_1 $ is the electric potential 
induced by $\mu _1 $ on $\Sigma _1 $, $V_2 $ is the electric potential 
induced by $\mu _2 $ inside $\Sigma _2 $ as well as on $\Sigma 
_1 $, and $V_{12} $ is the electric 
potential induced on $\Sigma _2 $ by $\mu _1 $. On the other hand, the total 
charge is

\begin{equation}
\label{eq34}
Q_1 = \int\limits_{{\bf{x}} \in \Sigma _1 } {\mu _1 \left( {\bf{x}} \right)d\Sigma _1 
\left( {\bf{x}} \right)} = C_1 V_1 
\end{equation}

\noindent
on $\Sigma _1 $ and

\begin{equation}
\label{eq35}
Q_2 = \int\limits_{{\bf{x}} \in \Sigma _2 } {\mu _2 \left( {\bf{x}} \right)d\Sigma _2 
\left( {\bf{x}} \right)} = C_2 V_2 
\end{equation}

\noindent
on $\Sigma _2 $. We mention that Eqs. (\ref{eq34}) and (\ref{eq35}) are valid only whenever $\Sigma _2 $ 
is one of the equipotential surfaces determined by an equilibrium charge distributed 
on $\Sigma _1 $. Equation (\ref{eq25}) provides the expressions of $C_1 $ and $C_2 $ 
that are the capacitances of $\Sigma _1 $ and $\Sigma _2 $, respectively. 
Moreover

\begin{equation}
\label{eq36}
Q_1 + Q_2 = 0
\end{equation}

\noindent
which is the Gauss law that is obtained from the second equation of (\ref{eq29}). 
Combining Eqs. (\ref{eq32}), (\ref{eq34}), (\ref{eq35}), and (\ref{eq36}) we 
obtain the capacitor capacity

\begin{equation}
\label{eq37}
C_{cond} =  \left( {\frac{1}{C_1 } - \frac{1}{C_2 }} \right)^{ - 1}.
\end{equation}
Particular examples in which Eq. (\ref{eq37}) is directly applicable is the capacitance of 
concentric spheres and of a coaxial cable. 
For a capacitor made of two concentric spheres the capacitance is 
$C_{sph\_cond}=\frac{4\pi \varepsilon R_1R_2}{R_2-R_1}$, where $R_1$ and $R_2$ are the 
radii of the two spheres with $R_2 > R_1$. Having in mind the capacitance of a sphere as $C_{sph}= 4 \pi \varepsilon R$, 
it is easy to check 
that $C_{sph\_cond}$ has the form given by Eq. (\ref{eq37}). Moreover, the capacitance of a capacitor 
made of two confocal spheroids obeys also (\ref{eq37}), with $C_1$ and $C_2$ as having analytic 
expressions that are given in the standard textbooks of classical electrodynamics \cite{Landau1984}.  

\begin{figure}[ht]
 \begin{center}
 \includegraphics {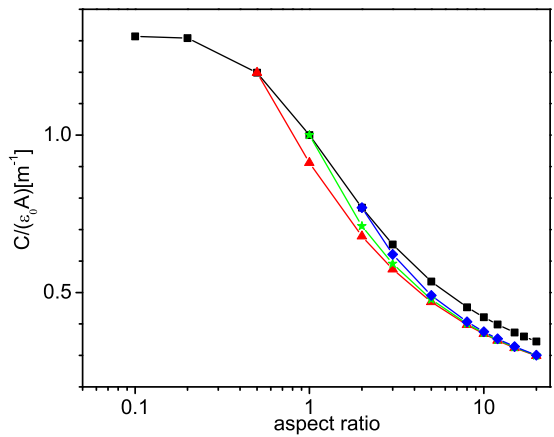}
 \end{center}
   \caption{\label{fig:2} (Color online) Calculated capacitance per unit area versus aspect ratio for spheroids (black squares) and for 
cylidrical rods with different end-cap geometries: oblate 
spheroid (red triangles), sphere (green stars), and prolate spheroid (blue diamonds). It is assumed that the 
largest cross-sectional radius is 1 m.}
\end{figure}

\section{A Numerical method. Discussion}

As we have previously discussed, capacitance can be determined in numerical simulators used to calculate other 
physical properties like the localized plasmon resonances in metallic nanoparticles \cite{Ouyang1989,Mayergoyz2005,Davis2009}. 
The method is an operator based BIE method that 
calculates numerically the eigenvalues $\chi_k$ and the eigenvectors $u_k$ and $v_k$ of $\hat M$ and $\hat M^\dag$, 
respectively. In order to have normed $u_k$ and $v_k$ one needs also to calculate the matrix elements of $\hat S$ \cite{Sandu2013}. 
The method belongs to the class of 
the spectral methods, where the functions of the basis set 
are defined globally rather than locally like in the FEM \cite{Boyd2001}. Details of the method for axisymmetric 
objects are given in Refs. \cite{Sandu2010,Sandu2011}. We have performed calculations for oblate and prolate spheroids as 
well as for cylindrical rods with different end-cap geometries: half of an oblate spheroid with $1/2$ aspect ratio, half of a sphere, and half of a 
prolate spheroid with an aspect ratio of $2$. 
To set the input data we have considered that the largest cross-sectional radius is 1 m for all the above geometries. The aspect ratio of any 
considered object is the ratio between the largest axial length and the largest cross-sectional diameter. The results are given in 
Fig. \ref{fig:2}, where the capacitance per unit area is plotted. For the same aspect ratio the rods have larger capacitances than the spheroids. Moreover, for aspect ratios 
greater than 5 the capacitance of the rods do not depend any longer on the end-cap geometry.

Our numerical calculations show a very good agreement with the analytical results for spheroidal shapes which can be found in 
Refs. \cite{Landau1984}. The relative error is at least $5 \times 10^{-5}$ with a relative small overhead of 25 functions in the basis and 
96 quadrature points. We have also checked these results against the more sophisticated 
Trefftz based FEM calculations provided by the multi-physics program ANSYS \cite{ansys}. The relative error in this case is around $10^{-2}$ 
(about $1\%$ and two order 
of magnitude greater than the BIE results). In ANSYS the capacitance results obtained with the Trefftz method are much better than the results obtained by 
using infinite elements. On the other hand, we compared the BIE calculations with ANSYS Trefftz calculations for the rods and the two 
calculations are apart by less than $1\%$. An example of rods with finite length is that of capped carbon nanotubes which are used as 
electrochemical double-layer capacitors (supercapacitors) for energy storage \cite{Huang2010}. The implementation of BIE for axisymmetric shapes has also shown to provide very accurate results of 
the depolarization factors which are related to the other eigenvalues of $\hat M$ and $\hat M^\dag$ \cite{Sandu2012}. 

A direct application of these calculations is the estimation of membrane capacitance in living cells. The shelled ellipsoidal model is 
one of the most common models of living cells in modeling dielectric spectroscopy 
experiments \cite{Asami1980}. In the ellipsoidal shelled model the shell designates the cell 
membrane that is bounded by two confocal ellipsoids and it is practically non-conductive in radiofrequency. Living cells accumulates positive/negative 
charge on the outer/inner surface of the 
membrane, giving rise to a resting potential across the membrane. In dielectric spectra the charge on the membrane is responsible for 
the $\alpha$-relaxation (usually below the frequency of 10 KHz), whilst the dielectric mismatch between the cell and the surrounding medium 
gives rise to $\beta$-relaxation in the MHz range of the radiofrequency spectrum \cite{Stoy1982}. The cell resting potential may be an input 
parameter in theoretical analysis of both $\alpha$- 
and $\beta$-relaxations, thus the membrane capacitance needs to be estimated \cite{Prodan1999, Prodan2008}. In the spherical model of living cells Prodan 
et al. \cite{Prodan2008} used the expression of a parallel-plane capacitor for the normalized capacitance (capacitance divided by area). 
It turns out that the formula of a parallel-plane capacitor is very close to that of a spherical living cell. The cell membrane is very thin with 
respect to the overall size of the cell, hence, for a spherical model of living cells, 
the membrane capacitance is approximated by 

\begin{equation}
\label{eq38}
C_{sph\_membr} \approx {\varepsilon A_2}/{d} = {C_{2}}/{(\delta-1)}, 
\end{equation}
where $d= R_2-R_1$, with $R_1$ and $R_2$ as the inner and outer membrane radius of areas $A_1$ and $A_2$, respectively. $C_{2}$ is 
capacitance of the outer sphere and $\delta =R_2/R_1>1$ obeys the condition $(\delta-1)\ll 1$. In the general case of arbitrarily shaped cells 
the parallel-plane capacitor formula cannot be used anymore although intuition may suggest otherwise. An analytical formula for the membrane 
capacitance of spheroidal living cells is provided by the combination of capacitance formula 
for spheroids \cite{Landau1984} and Eq. \ref{eq37}.

\begin{figure}[htp]
  \begin{center}
   \subfigure {\label{fig:3a} \includegraphics {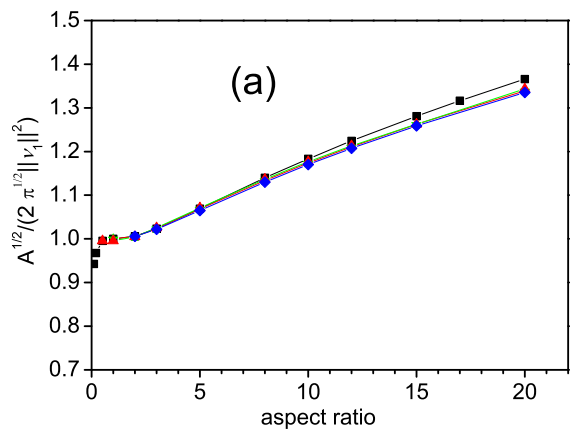}}  \\
    \subfigure {\label{fig:3b} \includegraphics {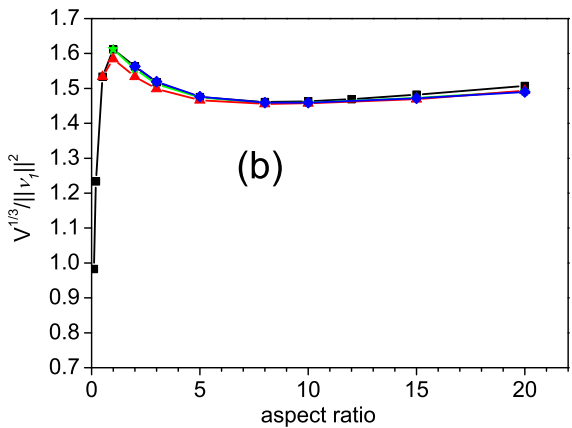}} 
  \end{center}
  \caption{(Color on-line) Scale-invariant shape factors versus aspect ratio. The shape factors are defined (a) by the area of the 
object and (b) by its volume. 
The spheroids are denoted by black squares and the cylidrical rods with different end-cap geometries are depicted by red triangles 
(oblate spheroid end-cap), by green stars (spherical end-cap), and by blue diamonds (prolate spheroid end-cap).}
  \label{fig:3}
\end{figure}

In Fig. \ref{fig:3} there are presented two shape factors that are related to capacitance and are scale-invariant. 
These shape factors can be straightforwardly utilized in approximate capacitance calculations for metallic object of various shapes. 
The first scale-invariant shape 
factor ($=A^{1/2}/(2 \pi^{1/2}||v_1||^2)$) is related to the area of the object 
(Fig. \ref{fig:3}a) and it shows a relative shape insensitivity for aspect ratios less than 5. This shape factor has 
been used in isoperimetric inequalities to estimates the capacitance of objects with shapes close to the spherical shape \cite{Chow1982}. In fact, Chow 
and Yovanovich noticed that this shape factor varies little over a wide range of shapes. 

The second scale-invariant shape factor related to capacitance is $V^{1/3}/||v_1||^2$. It is defined by using the volume of the object and shows 
shape insensitivity for long structures (Fig. \ref{fig:3}b). 
Thus for an aspect ratio greater than 5 the rods 
and the spheroids have almost the same volume related shape factors, which also vary little not only with the shape but also with respect to the aspect ratio. 
We note here that quantum capacitance of molecular nanowires is scaling with $V^{1/3}$ rather than with the length of the 
nanowire \cite{Ellenbogen2007}, thus for long structures 
the volume rather than the area of the object plays a greater role in determining the capacitance of an object.

\section{Summary and Conclusions}
We have presented a very compact formula of the capacitance for an arbitrarily shaped metallic object. 
Capacitance can be calculated with the help of the Neumann-Poincar\'{e} operator, which together with 
its adjoint plays a central role in the resolution of elliptic partial differential equations like the Laplace equation. 
The formula of capacitance is stated simply as follows. The capacitance is direct proportional to the dielectric permittivity of the embedding medium and 
to the area of the object, and inverse proportional to the squared norm of the eigenfunction of the Neumann-Poincar\'{e} operator 
with the largest eigenvalue. 
The operator approach on capacitance permits an elegant proof of the Thomson theorem which says that the electrostatic energy accumulated on a metallic body reaches its minimum when the charge distribution on the body is in equilibrium.
In addition, the capacitance and the charge distribution on a metallic object is a byproduct obtained in BIE numerical simulators of 
localized plasmon resonances. 

We have shown also how the capacitance of a capacitor can be related to the individual capacitance of each surface of the capacitor. Thus, when the 
outer surface of the capacitor is an equipotential surface generated by the charging of the inner surface then the capacitor behaves like a series 
capacitor with the total capacitance as being the capacitance of the inner surface in series with the opposite (negative) 
capacitance of the outer surface of 
the capacitor. 
These results have been used in the analysis of membrane capacitance in living cells, where the 
parallel-plane capacitor model works 
for spherical cells but is not appropriate for membrane capacitance of arbitrarily shaped living cells.

The capacitance formula allows us to define scale-invariant shape factors that can be used in the 
approximate calculation of capacitance. We have analyzed two scale-invariant shape factors. One of 
the shape factors employs the 
volume of the object and is 
more suitable for long shapes like rods or wires. Alternatively, the other shape factor, which is defined in terms of the object area, 
is appropriate for objects with shapes close to a sphere. 

\section{Acknowledgments}
This work was supported by a grant of the Romanian National Authority for 
Scientific Research, CNCS -- UEFISCDI, project number PNII-ID-PCCE-2011 
-2-0069.







\end{document}